\documentstyle[aps,preprint,tighten,floats]{revtex}

\newcommand{\lsim}{\mathrel{\mathop{\kern 0pt \rlap
  {\raise.2ex\hbox{$<$}}}
  \lower.9ex\hbox{\kern-.190em $\sim$}}}
\newcommand{\gsim}{\mathrel{\mathop{\kern 0pt \rlap
  {\raise.2ex\hbox{$>$}}}
  \lower.9ex\hbox{\kern-.190em $\sim$}}}

\newcommand{\beq}    {\begin{equation}}
\newcommand{\eeq}    {\end{equation}}
\newcommand{\beqarr} {\begin{eqnarray}}
\newcommand{\eeqarr} {\end{eqnarray}}
\newcommand{\barr}   {\begin{array}}
\newcommand{\earr}   {\end{array}}

\begin{document}
\preprint{
\begin{tabular}{r}
DFTT 75/96 \\
GEF--Th--19/96\\
ROM2F-96/63
\\
December 1996
\end{tabular}
}

\title{Exploring the supersymmetric parameter space by direct search for 
WIMPs}

\author{\bf A. Bottino$^{\mbox{a}}$
\footnote
{E--mail: bottino@to.infn.it,donato@to.infn.it,
mignola@to.infn.it,scopel@ge.infn.it,\\belli@roma2.infn.it,
incicchitti@vaxrom.roma1.infn.it},
F. Donato$^{\mbox{a}}$, G.
Mignola$^{\mbox{a,b}}$, S. Scopel$^{\mbox{c}}$
\vspace{6mm}}

\address{
\begin{tabular}{c}
$^{\mbox{a}}$
Dipartimento di Fisica Teorica, Universit\`a di Torino and \\
INFN, Sezione di Torino, Via P. Giuria 1, 10125 Turin, Italy
\\
$^{\mbox{b}}$
LAPP - Annecy Le Vieux, Chemin de Bellevue BP 110-F-74941 Annecy-Le-Vieux,
France
\\
$^{\mbox{c}}$
Dipartimento di Fisica, Universit\`a di Genova and\\
INFN, Sezione di Genova, Via Dodecaneso 33, 16146 Genoa, Italy
\end{tabular}}

\vspace{8mm}
\author{\bf
P. Belli$^{\mbox{d}}$, A. Incicchitti$^{\mbox{e}}$}
\address{
\begin{tabular}{c}
\\
$^{\mbox{d}}$
Dipartimento di Fisica, Universit\`a di Roma "Tor Vergata"\\
and INFN, sez Roma II, I-00173 Rome, Italy
\\
$^{\mbox{e}}$
Dipartimento di Fisica, Universit\`a di Roma "La Sapienza"\\
and INFN, sez. Roma, I-00165 Rome, Italy
\\
\end{tabular}}

\maketitle

\begin{abstract}
It is shown that the new data of the 
DAMA/NaI detector, combined with  recent calculations 
of nuclear structure functions, 
provide the most stringent upper bounds on the WIMP-nucleon
cross-section both for spin-dependent and for scalar interactions. 
It is proved that, for the first time, data from an experiment of 
direct search for WIMPs allow the  exploration of a region of the 
supersymmetric parameter space not yet fully investigated at accelerators. 
\end{abstract}  
\noindent
\newpage

\section{Introduction}

  An experiment for direct detection of Weakly Interacting Massive Particle 
  (WIMP),
   which employs a large-mass, low-background NaI(Tl) detector at the Gran Sasso
  Laboratory \cite{dama}, has recently allowed a substantial improvement of the
  upper limit on the WIMP-nucleon elastic scattering cross-section.  A high
  exposure of 4123 Kg day, a strong reduction of the residual internal
  contaminants and a statistical pulse shape discrimination have made this
  result possible. 

In this letter we examine the impact that this new measurement has in 
terms of
relic neutralinos, and more generally in terms of an exploration of the 
supersymmetric parameter space. 
First, in Sect.2  we discuss how upper bounds on (scalar and
spin-dependent) WIMP-nucleon cross-sections are extracted from the 
experimental counting rates. Whereas the procedure is straightforward for the
scalar case, for the spin-dependent case appropriate specifications on the 
interaction mechanism and realistic nuclear-physics calculations are needed. 
We show how the new data from
the DAMA/NaI-detector provide the most stringent upper bounds both to scalar 
and 
to spin-dependent WIMP-nucleon cross-sections (except for a very narrow 
range of the  WIMP mass around 15 GeV). This remarkable result derives from the 
significant improvement in sentivity in the experiment of Ref. \cite{dama} 
and, for the spin-dependent part, also 
from new theoretical calculations of  nuclear effects \cite{ressel,ressel1}. 
Then, in Sect.3 we proceed to a theoretical analysis of the implications
that the new experimental data have for relic neutralinos, and we show that, 
for the first time, data from an experiment of direct 
search for WIMPs allow the  exploration of some regions of the supersymmetric 
parameter space, whose investigation at accelerators is still under way.

\section{Rates for direct detection}

The experimental activity in the direct search for relic particles has been
very intense in recent years 
\cite{Ge,got,efo,fush,beck,twin,cosme,boulby,belli,rome,qxe,milano,sarsa,ei}, 
with quite significant improvements 
in various technical aspects that have drawn the experimental sensitivity 
nearer 
and nearer to the one required to match the predictions of 
realistic theoretical models. 
It is well known that the only safe way of
discriminating a real signal for WIMPs from the background is to measure 
a modulation effect \cite{freese}. 
Some preliminary measurements \cite{belli,sarsa,ei} have already 
been performed, 
but the search for modulation will
have to be pursued with increased sensitivity in order to reach the level 
required by a number of specific dark matter candidates 
(such as the neutralino). 
Since the current 
counting rates are still expected to be dominated by background effects, only upper bounds on
WIMP-nucleus cross-sections can be derived at present. 
Some of the considerations  presented in this section apply to a generic
WIMP, however, since we will concentrate on the neutralino ($\chi$) later on, 
when we compare experimental bounds with theoretical evaluations, we start from
the very beginning with a formalism which applies directly to this favorite
particle candidate for cold dark matter.

The neutralino is defined as the linear superposition 

\begin{equation}
\chi = N_1 \tilde \gamma + N_2 \tilde Z + N_3 \tilde H_1^{\circ}  
+N_4 \tilde H_2^{\circ}, 
\label{eq:neu}
\end{equation}

\noindent
of lowest mass. Here 
$\tilde \gamma, \tilde Z$ are the photino and zino states and 
$\tilde H_1^{\circ}$, $\tilde H_2^{\circ}$ are 
 the higgsino fields, 
supersymmetric partners of the Higgs
fields $H_1^{\circ}$, $H_2^{\circ}$. The theoretical framework used in this
paper for
the supersymmetric model will be presented in Sect.3.

We write the differential event rate for elastic neutralino-nucleus 
scattering as 

\beq
\frac {dR}{dE_R}=N_{T}\frac{\rho_{\chi}}{m_{\chi}}
                    \int^{v_{max}}_{v_{min}(E_{R})}\,dv\,f(v)\,v
                    \frac{d\sigma}{dE_{R}}(v,E_{R}) 
\label{eq:rate}.
\eeq
$N_{T}$ is the number density of the detector nuclei,
$\rho_{\chi}$/$m_{\chi}$ is the local 
(solar neighbourhood) 
number density of neutralinos,
$f(v)$ is the velocity distribution of neutralinos 
in the Galaxy (assumed to be Maxwellian in the Galactic rest frame) evaluated 
in the Earth's rest frame,
$d\sigma$/$dE_{R}$ is the elastic differential neutralino-nucleus 
cross section,
$E_{R}=m_{red}^{2}v^{2}(1-\cos{\theta^{*}})/m_N$ is the recoil energy 
($\theta^{*}$ is the 
scattering angle in the neutralino-nucleus center of mass frame),
$m_{red}$ is the neutralino-nucleus reduced mass and $m_N$ is the 
nuclear mass. 
Eq.(\ref{eq:rate}) is written for a monoatomic material; its generalization to
more complex materials is straightforward.

\subsection{Astrophysical and cosmological parameters}

Many astrophysical
quantities, affected by large uncertainties, enter 
in the evaluation of the differential event rate. This is the case for 
the 
neutralino 
 r.m.s. velocity and for its escape velocity, whose typical values  are: 
$v_{r.m.s.} = 270 \pm 24~ {\rm km \cdot s^{-1}}$
\cite{ffg}, 
$v_{esc} = 650 \pm 200~{\rm km \cdot s^{-1}}$
\cite{lt} and 
for the velocity of the Sun around the galactic centre 
($v_{\odot} = 232 \pm 20~ {\rm km \cdot s^{-1}}$\cite{ffg}).  

Also a  large 
uncertainty  concerns the value of the local dark 
matter density $\rho_l$. A recent determination of $\rho_l$, based on 
a flattened dark matter distribution and microlensing data, gives the range 
$\rho_l = 0.51_{-0.17}^{+0.21}~{\rm GeV \cdot cm^{-3}}$  \cite{turner}. In
particular the 
central value turns out to be significantly larger than  the one previously 
determined, for instance,  in Ref.\cite{flores}: 
$\rho_l = 0.3 \pm 0.1~{\rm GeV \cdot cm^{-3}}$. Furthermore, for any specific
value for the local density of  the {\it total} dark 
matter $\rho_l$, one has to assign a value to the 
{\it neutralino} local density 
$\rho_{\chi}$. To determine the value of $\rho_\chi$ to be
used in Eq.(\ref{eq:rate}), we adopt  the following rescaling recipe 
\cite{gaisser}: for each point of the parameter
space, we take into account the relevant value of the cosmological neutralino
relic density. When $\Omega_\chi h^2$ is larger than a minimal
$(\Omega h^2)_{min}$, compatible with observational data and with large-scale 
structure calculations, we simply put $\rho_\chi=\rho_l$.
When $\Omega_\chi h^2$ turns out  to be less than $(\Omega h^2)_{min}$, 
and then the neutralino may only provide a fractional contribution
${\Omega_\chi h^2 / (\Omega h^2)_{min}} \equiv  \xi$
 to $\Omega h^2$, we take $\rho_\chi = \rho_l \xi$.
The value to be assigned to $(\Omega h^2)_{min}$ is
somewhat arbitrary, in the range 
$0.03 \lsim (\Omega h^2)_{min} \lsim 0.3$. 

In Table I we give the two sets of astrophysical and cosmological parameters
that we use in this letter. Set I corresponds essentially to the central 
values for the
various parameters, whereas set II corresponds to those values of the
parameters, which, within the 
relevant allowed ranges,  provide the lowest detection rates 
(once the supersymmetric variables are fixed).

\begin{table}
\caption{Values of the astrophysical and cosmological parameters 
relevant to direct detection rates. 
$V_{r.m.s.}$ denotes the root mean square velocity of the neutralino Maxwellian
velocity distribution in the halo; $V_{esc}$ is the neutralino escape velocity
and $V_{\odot}$ is the velocity of the Sun around the galactic centre; 
$\rho_{loc}$ denotes the local dark matter density and $(\Omega h^2)_{min}$ the
minimal value of $\Omega h^2$. The values of set I are the median values of
the various parameters, the values of set II are the extreme values of the
parameters which, within the physical ranges, provide the lowest estimates of 
the direct rates
(once the supersymmetric parameters are fixed).
}
\begin{center}
\begin{tabular}{|c|c|c|}   \hline
 &  Set I &  Set II \\ \hline
$V_{r.m.s}(\rm km \cdot s^{-1}$) & 270 & 245 \\ \hline
$V_{esc}(\rm km \cdot s^{-1}$)   & 650 & 450 \\ \hline
$V_{\odot}(\rm km \cdot s^{-1}$) & 232 & 212 \\ \hline
$\rho_{loc}(\rm GeV \cdot cm^{-3}$) & 0.5 & 0.2 \\ \hline
$(\Omega h^2)_{min}$            & 0.03 & 0.3 \\ \hline
\end{tabular}
\end{center}
\end{table}

\subsection{Cross-sections}

The differential cross-section $\frac{d\sigma}{dE_{R}}(v,E_{R})$  
in Eq.(\ref{eq:rate}) may
conveniently be splitted into a coherent part and a spin-dependent 
one \cite{gw}

\beq
\frac{d \sigma}{d E_R}=\left(\frac{d \sigma}{d E_R}\right)_{C}+
\left(\frac{d \sigma}{d E_R}\right)_{SD}.
\eeq

 When the WIMP is the neutralino, the coherent cross-section, 
$(\frac{d\sigma}{dE_{R}})_C$, originates from Higgs-exchange \cite{barb} and
 squark-exchange \cite{griest} in the neutralino-nucleus scattering; the 
spin-dependent cross-section 
$(\frac{d\sigma}{dE_{R}})_{SD}$ 
is due to Z-exchange and to 
squark-exchange \cite{griest}. The links among supersymmetric and nuclear 
degrees of freedom have quite different features in the coherent and in the
spin-dependent cross-sections. Thus, let us consider the two cases separately. 

\subsubsection{Coherent $\chi$--nucleus cross-section}

In the case of the  coherent cross-section,  supersymmetric and nuclear 
degrees of freedom decouple, thus $(\frac{d\sigma}{dE_{R}})_C$ may 
be written as

\beq
\left(\frac{d \sigma}{d E_R}\right)_{C}=
\frac{\sigma_{C}^0}{E_R^{max}} F^2(q)
\label{eq:dsigma_dq_coer}
\eeq

\noindent
where $\sigma_{C}^0$ is the total coherent cross-section, conventionally 
 integrated from 
zero up to $E_R^{max}$ $=$ $2 m_{red}^2 v^2/m_N$, 
$q^2 \equiv\vec{q}\;^2=2m_NE_R$ is the squared
three--momentum transfer,
and $F_C(q)$ is a nuclear form factor 
defined as the Fourier transform of the nuclear matter distribution. 
Except for very light nuclei, which have a form factor 
$F_C$ of  exponential or  
gaussian type, 
a universal parametrization for $F_C(q)$ is provided by \cite{helm} 

\begin{equation}
F(q)=3 \frac {j_1(qr_0)}  {qr_0} e^{-\frac {1}{2} s^2 q^2},
\label{eq:cff}
\end{equation}

\noindent 
where $j_1(qr_0)$ is the spherical Bessel function of index 1,
$s \simeq 1~ \rm fm$ is the
thickness parameter for the 
the nuclear surface, $r_0 = (R^2-5s^2)^{1/2}$ and $R=1.2~A^{1/3}$ fm 
($A$ is the atomic number). 

\noindent
The cross-section 
$\sigma_{C}^0$ may be written as 

\beqarr
\sigma_{C}^0 &=& \frac {8 G_F^2} {\pi} M_Z^2 \zeta^2 m_{red}^2 A^2,
\label{eq:co}
\eeqarr

\noindent
where the quantity $\zeta$ depends on the $\chi - quark$ couplings mediated by 
Higgs particles and squarks \cite{barb,griest}. The full expression for $\zeta$ 
together with the values used here for the relevant parameters are reported in
Ref.\cite{scopel}. It is worth recalling that the simple dependence of 
$\sigma_C$ on $A^2$ is due to the equality (within a very good accuracy) of the
$\chi$-neutron and the $\chi$-proton couplings. 

From the factorization property inherent in the structure of Eq.(\ref{eq:co})
it follows that $\sigma_{C}^0$ may be immediately converted into an
equivalent {\it $\chi-$ nucleon scalar} cross section

\beq
\sigma^{(nucleon)}_{scalar}=\frac{1+m^{2}_{\chi}/m^2_N}{1+m^{2}_{\chi}/m^2_p}
\frac{\sigma^0_{C}}{A^{2}}
\label{eq:nucleon}
\eeq

\noindent
where $m_p$ denotes the proton mass. 
 Eqs.(\ref{eq:dsigma_dq_coer}-\ref{eq:nucleon}) enable a straightforward 
conversion of  any upper limit on 
$\frac{d \sigma}{d E_R}$ into an upper bound on 
$\sigma^{(nucleon)}_{scalar}$. This procedure allows a direct comparison among the
results of various experiments. It is obvious that this method applies also 
to any
other WIMP which couples with equal strength to neutrons and to protons. 

In Fig.1 we present a summary of the
current upper bounds, obtained by the method previously described. All
experimental data  have been treated with the same algorithm and with the same 
set of astrophysical parameters (set I above; ($\Omega h^2)_{min}$ is not 
relevant here). 
We notice that the upper limit derived from the DAMA/NaI
experiment provides the most stringent upper bound
on the scalar neutralino-nucleon 
cross-section with an
improvement of a factor 3--5 over the limits presently placed by Ge-detectors. 
The only exception is represented by  a very narrow range in the neutralino
mass 
around $m_{\chi} \simeq 15$ GeV. However, except for peculiar values of the
other supersymmetric parameters, this range in $m_{\chi}$ is below the current 
lower bound on the neutralino mass \cite{zeit,efos}.

\subsubsection{Spin-dependent $\chi$--nucleus cross-section}

Let us turn now to the cross-section due to spin-dependent interactions. 
Using standard notations, 
we write the differential cross-section $\frac{d \sigma}{d E_R}$ for 
neutralino-nucleus scattering as

\beqarr
\left(\frac{d \sigma}{d E_R}\right)_{SD}=
\frac{16 G_F^2}{v^2} m_N
\frac{S(q)}{2J+1} . 
\label{eq:dsigma_dq_spin}
\eeqarr

\noindent 
where $J$ is the nuclear spin and $S(q)$ is given by:
\beq
S(q)=a_0^2 S_{00}(q)+a_1^2 S_{11}(q)+a_0 a_1 S_{01}(q)
\eeq

The functions $S_{ij}, (i,j = 0,1)$ are rather complicated functions 
which  depend on the nuclear spin structure 
and on the neutralino composition; their definitions may be found in 
Ref.\cite{epv}.
The coefficients $a_0$ and $a_1$ are the isoscalar and isovector projections of the 
neutralino-neutron and neutralino-proton couplings $a_n, a_p$, respectively, 
{\it i.e.}
 $a_0$=$a_p$+$a_n$, $a_1$=$a_p$--$a_n$.
The full expressions for $a_n, a_p$ are reported, for instance, in Refs.
\cite{scopel,jkg}. In the case of the $\chi$-nucleon interaction mediated by 
Z-exchange one has 

\beqarr
&a_{p,n}&=-\frac{1}{2} (N_3^2-N_4^2) (\sum_{q=u,d,s} T_{3q} \Delta q)_{p,n} 
\label{eq:aa}
\eeqarr

\noindent
where $T_{3q}$ denotes the third component of the quark weak isospin, 
$\Delta q$ is the fractional spin carried by the quark q in the appropriate
nucleon and $N_3, N_4$ are the Higgsino coefficients in the  expression 
for  $\chi$ (see Eq.(\ref{eq:neu})).

The cross-section of Eq.(\ref{eq:dsigma_dq_spin}) may be conveniently rewritten as 
\cite{rab}  

\beqarr
\left(\frac{d \sigma}{d E_R}\right)_{SD}=
\frac{16 G_F^2}{v^2 \pi} m_{N} \Lambda^{2}J(J+1)
\frac{S(q)}{S(0)}, 
\eeqarr

\noindent 
where the static value of $S(0)$ has been expressed as 

\beq
S(0)=\frac{2J+1}{\pi}\,\Lambda^{2}J(J+1)
\eeq
 
with 

\beq
\Lambda=\frac{a_p <\vec{S}_p>+a_n <\vec{S}_n>}{J}.
\eeq

\noindent 
$<\vec{S}_p>$ and $<\vec{S}_n>$ are the contributions of the total proton 
and neutron spins in the nucleus. Another commonly used notation is 
$\lambda_{p(n)} = \Lambda/a_{p(n)}$. 

It is important to recall that both $\Lambda$ and $S(q)/S(0)$ depend on nuclear
properties as well as on the neutralino composition. Thus, in spin-dependent
interactions nuclear degrees of freedom and supersymmetric degrees of freedom 
do not decouple in the most general case. This feature does not allow an
extraction of a universal ({\it i.e.}  independent of the neutralino
composition) spin-dependent $\chi-$nucleon cross-section in 
the same way as was done in the coherent case. 

An exception is provided by
spin-dependent interactions mediated by Z-exchange only. This property is
explicitely exhibited by the expressions of Eq.(\ref{eq:aa}) for $a_p, a_n$, 
which imply 

\beqarr
[a_{n}/a_{p}]_{Z}=\left(\sum_{q}T_{3q}\Delta q\right)_{n}/
\left(\sum_{q}T_{3q}\Delta q\right)_{p}
\eeqarr

In order to compare experimental data due to various detectors in a way similar
to the one previously employed for the coherent interactions, in the rest of
this section we expand the formalism in the case of a pure Z-exchange
contribution to the spin-dependent scattering. Under this assumption, 
the spin-dependent cross-section may be cast
into the form

\beqarr
\left(\frac{d \sigma}{d E_R}\right)_{SD}=
\frac{4 G_F^2}{\pi v^2 } m_{N} (N_{3}^{2}-N_{4}^{2})^{2}
\left(\sum_{q}T_{3q}\Delta q\right)
_{p}^{2}\,\lambda_{p,Z}^{2}J(J+1) \,F^2_{SD}(q)
\label{eq:sig}
\eeqarr
   
where $\lambda_{p,Z}$ is defined as 

\beqarr
\lambda_{p,Z}\equiv\left[\frac{\Lambda}{a_p}\right]_{a_n/a_p=[a_n/a_p]_Z}=
\frac{1}{J}\left[<\vec{S}_p>+\frac{a_n}{a_p} <\vec{S}_n>\right]_{a_n/a_p=[a_n/a_p]_Z}
\eeqarr

and 

\beq
F^2_{SD}(q)\equiv\left[\frac{S(q)}{S(0)}\right]
_{a_n/a_p=[a_n/a_p]_Z}
\eeq

Thus we may further define a total point-like spin-dependent cross-section
(conventionally integrated from zero up to $(E_R)_{max}$) as

\beqarr
\sigma^0_{SD}
& = & (E_R)_{max}\left(\frac{d\sigma}{d E_R}\right)_{SD}
\frac{1}{F^2_{SD}(q)} \nonumber \\
& = & \frac{8 G_F^2}{\pi} m_{red}^2 (N_{3}^{2}-N_{4}^{2})^{2}
\left(\sum_{q}T_{3q}\Delta q\right)
_{p}^{2}\,\lambda_{p,Z}^{2}J(J+1),  
\eeqarr

\noindent
in terms of which we may obtain an  equivalent $\chi-$proton 
spin-dependent cross-section 

\beq
\sigma^{(proton)}_{SD}=\frac{1+m^{2}_{\chi}/m^2_N}{1+m^{2}_{\chi}/m^2_p}
\frac{3}{4}\frac{\sigma^0_{SD}}{\lambda_{p,Z}^2 J(J+1)}
\label{eq:sigg}
\eeq

Therefore, under the assumption of a pure (or dominant) Z-exchange, 
Eqs.(\ref{eq:sig}-\ref{eq:sigg}) allow the conversion of any experimental 
upper limit on 
$\frac{d \sigma}{d E_R}$ into an upper bound on 
$\sigma^{(proton)}_{SD}$, 
in much the same way as for the coherent case. However, still an important 
difference between the two 
cases concerns the features of the nuclear properties. Whereas in the 
coherent case these 
are simply taken into account by the form factor $F_C$ of  the universal 
analytic form of Eq.(\ref{eq:cff}), in the spin-dependent one 
the nuclear properties have to be calculated for each nucleus with an 
appropriate nuclear 
model. In our analysis of the spin-dependent cross-sections 
we have employed the most recent nuclear-physics calculations for the 
structure functions $S_{ij}(q)$ for the nuclei of interest. 
 From the previous formulae 
it is clear that from these $S_{ij}$'s both the static value 
$\lambda$ and the form factor $F_{SD}$ are readily derived. 

The nuclei with non-vanishing spin,  relevant for our analysis, 
are $^{23}$Na, $^{73}$Ge, $^{127}$I, $^{129}$Xe. 
For these nuclei we give in Table   II the values for 
$\lambda_{p,Z}^2\;J(J+1)$ 
used in our analysis \cite{ressel,ressel1,dep}, compared with some of the 
previous determinations \cite{ef,ikm}. 
We have set  
$a_n/a_p = [a_n/a_p]_Z$, and we have assigned to this ratio the value  
$[a_n/a_p]_Z = -0.85$, which is obtained from Eq.(\ref{eq:aa}) using for the 
$\Delta q$'s the estimates of Ref.\cite{ek}.

\begin{table}
\caption{ Values of the nuclear spin-dependent static properties. 
In the first column are listed the nuclei with non-vanishing spin considered in
the present paper. Their nuclear spins are given in the second column. 
In the third, fourth and fifth columns are listed the values 
of $<\vec{S}_p>$, $<\vec{S}_n>$ and $\lambda_{p,Z}^2 J(J+1)$ derived 
from the following references: Ref.\protect\cite{ressel1} for $^{23}$Na and $^{129}$Xe, 
Ref.\protect\cite{dep} for $^{73}$Ge, and Ref.
\protect\cite{ressel,ressel1} for $^{127}$I. 
In the sixth column is given the ratio 
$(\lambda_{p(n),Z}/\lambda_{OG})^2$: the numerator 
stands for the $\lambda$ 
factor for proton or for neutron, according to whether protons or neutrons are 
the odd species in the relevant nucleus ($\lambda_{p,Z}$ for 
 $^{23}$Na and  $^{127}$I, and  
$\lambda_{n,Z}$ for $^{73}$Ge and 
 $^{129}$Xe), and  its value is obtained from 
the Refs.\protect\cite{ressel,ressel1,dep}; the denominator stands for previous 
evaluations with the odd-group approximation for $^{23}$Na, 
$^{73}$Ge and $^{129}$Xe \protect\cite{ef} and with the interacting boson 
fermion model
for $^{127}$I \protect\cite{ikm}. 
(B) and (N) denote that the nuclear-physics calculations of 
Refs.\protect\cite{ressel,ressel1} were performed with the Bonn A or with 
the Nijmegen II nucleon-nucleon potential, respectively.
} 
\begin{center}
\begin{tabular}{|c|c|c|c|c|c|} \hline 
 nucleus & J & $<\vec{S}_p>$ & $<\vec{S}_n>$ &
  $\lambda^2_{p,Z}J(J+1)$ & $\lambda^2_{p(n),Z}/\lambda^2_{OG}$
 \\ \hline \hline
${}^{23} \rm{Na}$ & 3/2 & 0.248 & 0.020 & 0.089 & 2.2 \\ \hline
${}^{73} \rm{Ge}$ & 9/2 & 0.030 & 0.378 & 0.103 & 2.2 \\ \hline
${}^{127} \rm{I}$ \it{(B)} & 5/2 & 0.309 & 0.075 & 0.084 & 3.6 \\ \hline
${}^{127} \rm{I}$ \it{(N)} & 5/2 &
 0.354 & 0.064 & 0.126 & 5.5 \\ \hline
${}^{129} \rm{Xe}$ \it{(B)} & 1/2 & 0.028 & 0.358 & 0.229 & 2.5 \\ \hline
${}^{129} \rm{Xe}$ \it{(N)} & 1/2 & 0.013 & 0.300 & 0.177 & 2.0 \\ \hline
\end{tabular}
\end{center}
\end{table}

It is noticeable that the 
new evaluations of the static properties provide 
 larger values for $\lambda$ as compared to the previous ones. 
We remark that  
these quantities are somewhat sensitive to the type of the nucleon-nucleon  
interactions used in the calculations.

The form factor $F_{SD}$ for the various nuclei has been calculated using the 
nuclear-physics
results of Refs.\cite{ressel,ressel1,dep}. 
The relevance of these finite-size effects depend 
on the range of the variable qR which is employed for the extraction of the 
upper bounds on the differential counting rates. This, in turn, depends (apart from
the nuclear size) on the other experimental features: energy threshold and 
level of the counting rates. 
For $^{23}$Na and $^{73}$Ge finite size effects are negligeable, since the 
relevant qR values are $0.27 \leq {\rm qR} \leq 0.55$ for $^{23}$Na and 
$0.75 \leq {\rm qR} \leq 1.4$ for $^{73}$Ge.
 On 
the contrary,  these effects have to be properly taken into account for 
$^{127}$I and  $^{129}$Xe, whose qR ranges are: 
$2.2 \leq {\rm qR} \leq 4.4$ for $^{127}$I and 
$1.6 \leq {\rm qR} \leq 3.5$ for  $^{129}$Xe.
In Fig.2 are shown the plots of the function 
$[S(q)/a_p^2]_Z \equiv [S(q)/a_p^2]_{a_n/a_p=[a_n/a_p]_Z}$ 
for these two nuclei,
 calculated from 
the structure functions $S_{ij}$'s of Refs.\cite{ressel,ressel1}
for the two different nucleon-nucleon interactions previously mentioned. 
Again we have set  
$a_n/a_p = [a_n/a_p]_Z = -0.85$. 
It is worth noticing two main features: i) the shape of $S(q)$ is not very 
sensitive to the form of the nucleon-nucleon interaction, except for 
$^{129}$Xe for ${\rm qR} \gsim 4$; a region  which is,  however, 
beyond the  qR-range employed in our 
extraction of the upper limits, ii) the function $S(q)$ for 
$^{127}$I shows a plateau for qR $\gsim$ 3, then in a region which 
is important in our analysis. 

 When we apply the procedure previously outlined ({\it i.e.} 
Eqs.(\ref{eq:sig}-\ref{eq:sigg})) to 
the counting rates of the experiments of 
Refs.\cite{dama,got,twin,cosme,boulby,rome} 
we finally obtain the 
results shown in Fig.3. 
For the quenching factor Q the following values have been used: Q = 0.30 
for $^{23}$Na \cite{dama,fush,boulby},  
Q = 0.25 for $^{73}$Ge \cite{got}, Q = 0.09 for $^{127}$I 
\cite{dama,fush,boulby} and 
Q = 0.65 for $^{129}$Xe \cite{belli,qxe}. 
From this figure it turns out that the data of 
the DAMA/NaI provide the most stringent upper bound 
also for the spin-dependent cross-section 
$\sigma_{SD}^{(proton)}$. It is noticeable 
that, because of the new higher value for $\lambda$ and of the property ii) 
 discussed above, the constraint from the I-nucleus turns out to be 
significantly stronger as compared to results of previous analyses.

\section{Evaluation of cross-sections in the MSSM}

The supersymmetric model employed in this paper is the so-called 
   Minimal Supersymmetric extension of the Standard Model 
(MSSM)\cite{Susy}, which 
 provides the most convenient phenomenological 
framework at the electroweak scale ($M_Z$) and does not assume too 
strong, arbitrary theoretical hypotheses (for instance, universality 
conditions for the soft scalar masses at the Grand Unified scale $M_{GUT})$. 
Various properties (relic abundances and detection rates) 
of relic neutralinos have been analyzed in the MSSM by 
a number of authors. Some of the most recent ones are given in 
Refs.\cite{noi,jkg,bg,bottino,mignola}.  

The MSSM is based on  the same gauge group as the Standard Model
and contains the supersymmetric extension of its particle content. 
  The
Higgs sector is modified as compared to that of the Standard
Model, because it requires
two Higgs doublets $H_1$ and $H_2$ in order to give mass both to down-- and
up--type quarks and to cancel anomalies. Thus the MSSM 
contains three neutral Higgs fields: a pseudoscalar and two scalar fields. 
The Higgs sector is specified at the tree level by
two independent parameters:
the mass of a physical Higgs field (we use the mass $m_A$ of the 
neutral pseudoscalar boson) and the ratio of the two vacuum
expectation values, usually defined as $\tan\beta=v_2/v_1 \equiv <H_2>/<H_1>$.
Other important ingredients of the model are, apart from the Yang-Mills 
Lagrangian, the superpotential, 
which contains  all the Yukawa interactions
and the Higgs-mixing term 
$\mu H_1 H_2$, and  the soft--breaking
Lagrangian, containing the trilinear and bilinear  breaking 
parameters and the soft gaugino and scalar masses.

The theoretical model 
contains an exceedingly large number of parameters, unless 
some  assumptions are introduced. It is customary to introduce the following 
conditions (at the $M_Z$ scale): i) all trilinear parameters are 
zero except those of the third family, which 
however have a common value A, ii) 
all squarks are taken as degenerate: $m_{\tilde q_i} \equiv m_0$, except 
for $m_{\tilde t}$, iii) all sleptons  are taken as degenerate 
with $m_{\tilde l_i} \equiv m_0$, iv) the gaugino masses are assumed to 
unify at $M_{GUT}$; 
thus in particular $M_1= (5/3) \tan^2 \theta_W M_2$. 

After these conditions are applied, the number of independent 
parameters  reduces to seven. We choose them to be: $M_2, \mu, \tan \beta, 
m_A, m_0, m_{\tilde t}$, A. 
In terms of these parameters, 
the supersymmetric space is further constrainted by
the following requirements: 1) all experimental bounds on Higgs, 
neutralino, chargino and
sfermion masses  are satisfied (taking into account also the new data from
LEP2 \cite{zeit,LEP2,delphi}), 2) the neutralino is the Lightest Supersymmetric
Particle (LSP) (i.e., regions where gluino or squarks or sleptons are 
LSP are excluded), 3) constraints due to the $b \rightarrow s + \gamma$ 
process  \cite{alam} are satisfied, 4)  the neutralino relic
abundance does not exceed the cosmological bound, i.e. 
$\Omega_{\chi}h^2 \leq 1$.

The results presented in this paper for the evaluations of 
neutralino-nucleon cross sections within MSSM 
have been obtained by varying the 
independent  parameters of the 
model  in the following ranges: 
$10\;\mbox{GeV} \leq M_2 \leq  500\;\mbox{GeV},\; 
10\;\mbox{GeV} \leq |\mu| \leq  500\;\mbox{GeV},\;
60\;\mbox{GeV} \leq m_A \leq  500\;\mbox{GeV},\; 
100\;\mbox{GeV} \leq m_0 \leq  500\;\mbox{GeV},\;
100\;\mbox{GeV} \leq m_{\tilde t}  \leq  500\;\mbox{GeV},\;
-3 \leq {\rm A} \leq +3,\;
1.1 \leq \tan \beta \leq 50$, and taking into account the experimental constraints 
previously mentioned. 
The ensuing explored range for $m_{\chi}$ is from its lower bound up to 
250 GeV.

Within the present theoretical framework we have evaluated the 
cross-sections $\sigma^{(nucleon)}_{scalar}$ and 
$\sigma^{(proton)}_{SD}$. 
The results for 
$\sigma^{(nucleon)}_{scalar}$, multiplied by the scaling factor $\xi$, 
 are shown in Fig.4. Allowed neutralino configurations stay inside the outer 
contour line. 
To provide more information, also the locations of configurations with the 
representative values: 60 GeV $\leq m_A \leq$ 70 GeV and 
$\tan \beta = 10, 50$ are displayed. 
The thick curve represents the experimental 
upper bound to $\sigma^{(nucleon)}_{scalar}$, deduced by using the set I 
of Table I; thus this curve is just a part of the one in Fig.1 
and is due to the new data from the DAMA/NaI experiment.  This figure shows 
that a large number of neutralino configurations are above the current 
experimental bound. It follows 
that the present experimental sensitivities have reached a level of great 
significance for the possibility of investigating  realistic 
supersymmetric models. However, this result cannot be used 
to disallow those configurations which provide  a 
$\xi \cdot \sigma^{(nucleon)}_{scalar}$ above the experimental limit, because of the 
uncertainties affecting the astrophysical parameters. 
If we assign  to these parameters their most conservative values, 
{\it i.e.} the values of set II of Table I, we obtain the results displayed in
Fig.5. This Figure shows that, even in the very conservative scenario
represented by set II, for the first time the boundary of the physical region
for neutralinos is reached by direct detection measurements. From our results
 it turns out that the neutralino configurations currently accessible by direct
detection are those pertaining to light Higgs
masses: $m_A \simeq 60$ GeV and large $\tan \beta$: 
$\tan \beta \simeq  40-50$. These same configurations are at present under
investigation at LEP2 (at $\sqrt s = 172$ GeV). 

A similar analysis, performed for the spin-dependent case, shows that the 
theoretical predictions within MSSM for 
$\sigma^{(proton)}_{SD}$ are still much below the current upper bound of
Fig.3 (by more than two orders of 
magnitude). This is expected, in view of the much reduced strength of
spin-dependent effects versus the coherent ones.

A few comments
are in order here. Investigations of fundamental particle-physics 
properties by non-accelerator experiments usually suffer from  a number of 
serious uncertainties.
This is  the case  for the problem under discussion, 
 as we have already recalled above, in connection with 
the astrophysical and cosmological parameters. 
For this reason, in our conclusions we have also 
conservatively taken into account the set II for these parameters. However, 
it is remarkable that, even in this unfavorable scenario, the present
sensitivity of experiments for direct 
WIMP search turns out to be at a level of great physical interest for 
supersymmetric relic particles. 
Significant improvements in
the determination of fundamental cosmological parameters (such as the 
average matter density and the Hubble constant) are expected in the near
future; these will allow  to draw  stronger conclusions from  investigations 
similar to the one presented here. 
It is also to be remarked that further substantial improvements 
in the sensitivities of experiments searching for WIMPs
are expected, thus the perspectives
for this kind of experimental investigation are very encouraging.

\section{Conclusions}

Using the results of 
  an experiment for direct detection of Weakly Interacting Massive Particle 
  (WIMP),
   which employs a large-mass, low-background NaI(Tl) detector at the Gran Sasso
  Laboratory, DAMA/NaI \cite{dama}, and new nuclear-physics evaluations of
  spin-dependent structure functions \cite{ressel,ressel1}, we have shown that 
these new experimental data provide the most stringent upper bounds 
both to scalar and 
to spin-dependent WIMP-nucleon cross-sections (except for a very narrow 
range for WIMP mass around 15 GeV). Applying these results to the most favorite
candidate for cold dark matter, the neutralino, we have proved that, for the first
time, data from an experiment of direct 
search for WIMPs allow the  exploration of a region of the supersymmetric 
parameter space not yet fully investigated at accelerators. 
This result has been obtained by using for the astrophysical and cosmological
parameters a very  conservative set of values. We may also conclude that the 
experiment of Ref.\cite{dama} can allow  an investigation of 
modulation effects with significant sensitivity.

{\bf Acknowledgments}

We thank Rita Bernabei for very interesting discussions. 
We also express our thanks to M.Ted Ressell for providing us with unpublished results
on spin-dependent nuclear structure functions.

\vfill
\eject

{\bf Figure Captions}

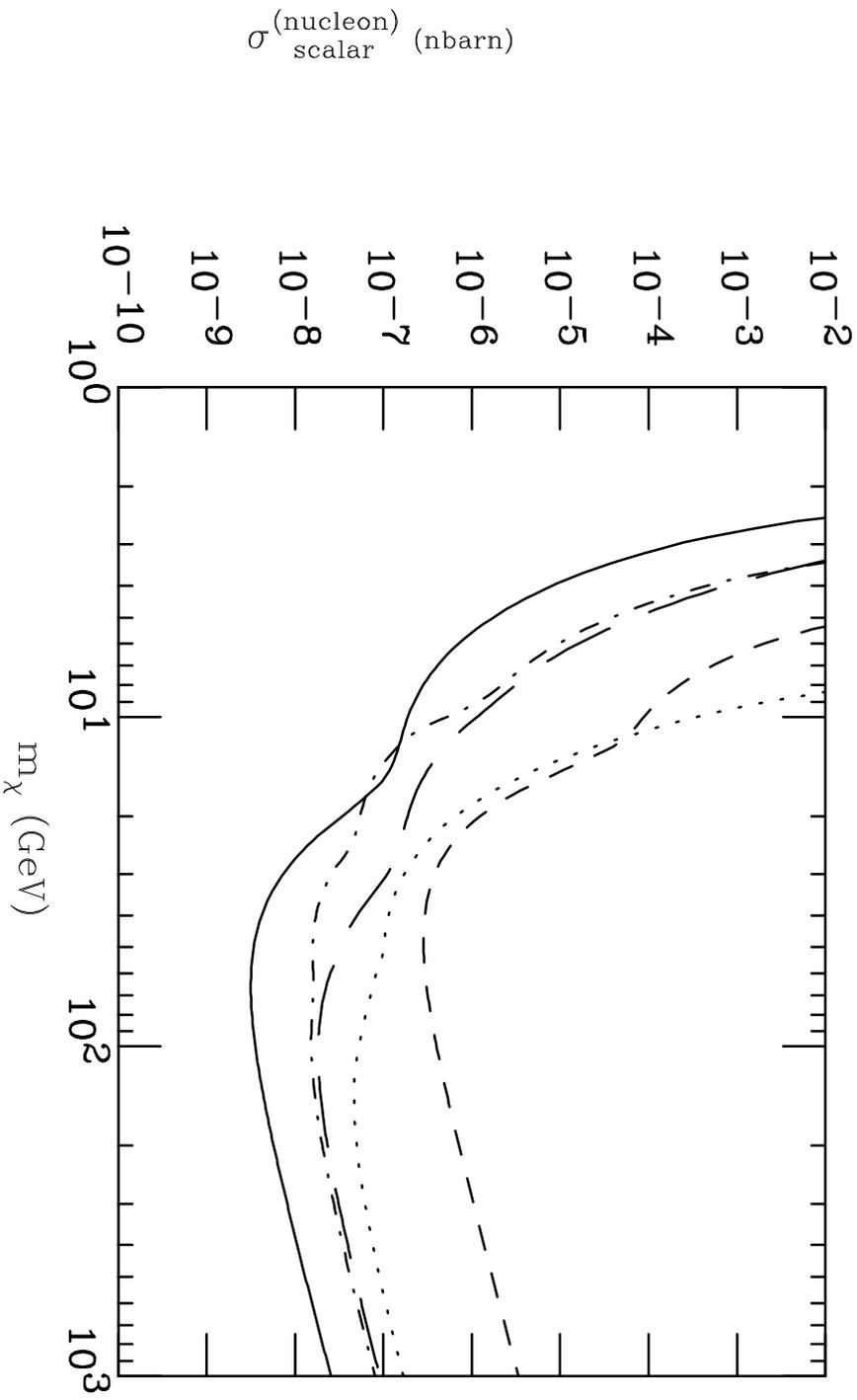
{\bf Figure 1} -- Upper bounds on $\sigma^{(nucleon)}_{scalar}$. The 
dot-dashed line denotes the limit obtained by combining  data of 
Ge-detectors; the branch for $m_{\chi} \lsim 10$ GeV is from results of 
Ref.\cite{cosme}, the one for 10 GeV$\lsim m_{\chi} \lsim$ 20 GeV is from 
results of Ref.\cite{got} and the one for $m_{\chi} \gsim$ 20 GeV is 
from data of Ref.\cite{twin}.  The solid and the long-dashed lines denote the 
bounds from the results of NaI-detectors: the solid line is from data of 
Ref.\cite{dama}, the long-dashed line is from data of Ref.\cite{boulby}. Dotted and 
short-dashed lines are from data of a Xe-detector \cite{rome} and of a
TeO$_2$-detector \cite{milano}, respectively. 

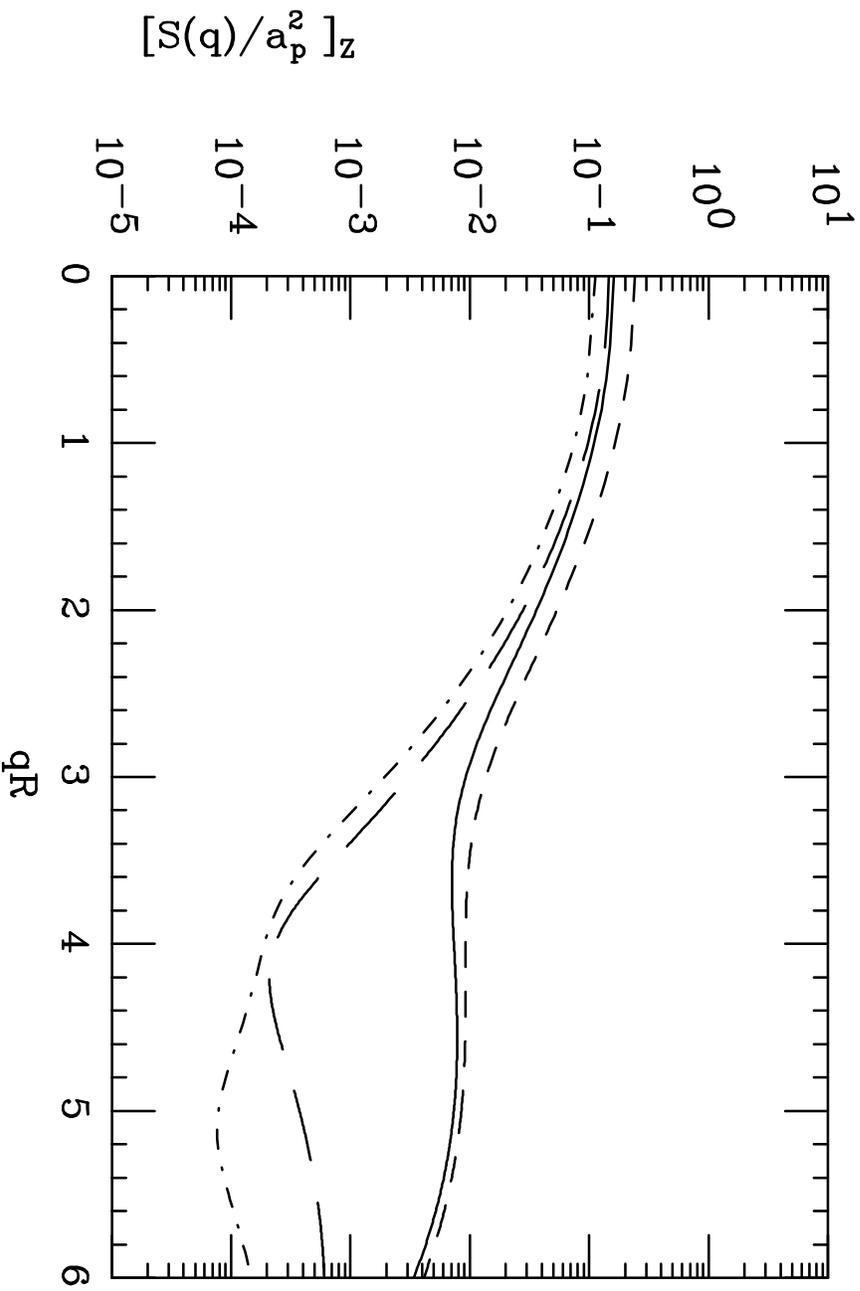
{\bf Figure 2} -- Structure function $[S(q)/a_p^2]_Z$, obtained 
from the results of Refs.\cite{ressel,ressel1}, versus  qR. 
The two lower curves refer to $^{129}$Xe 
 with the Bonn A (long-dashed line) and  with 
the Nijmegen II (dot-dashed line) nucleon-nucleon potentials.
The two upper curves refer to  $^{127}$I 
 with the Bonn A (short-dashed line) and  with 
the Nijmegen II (solid line) nucleon-nucleon potentials.

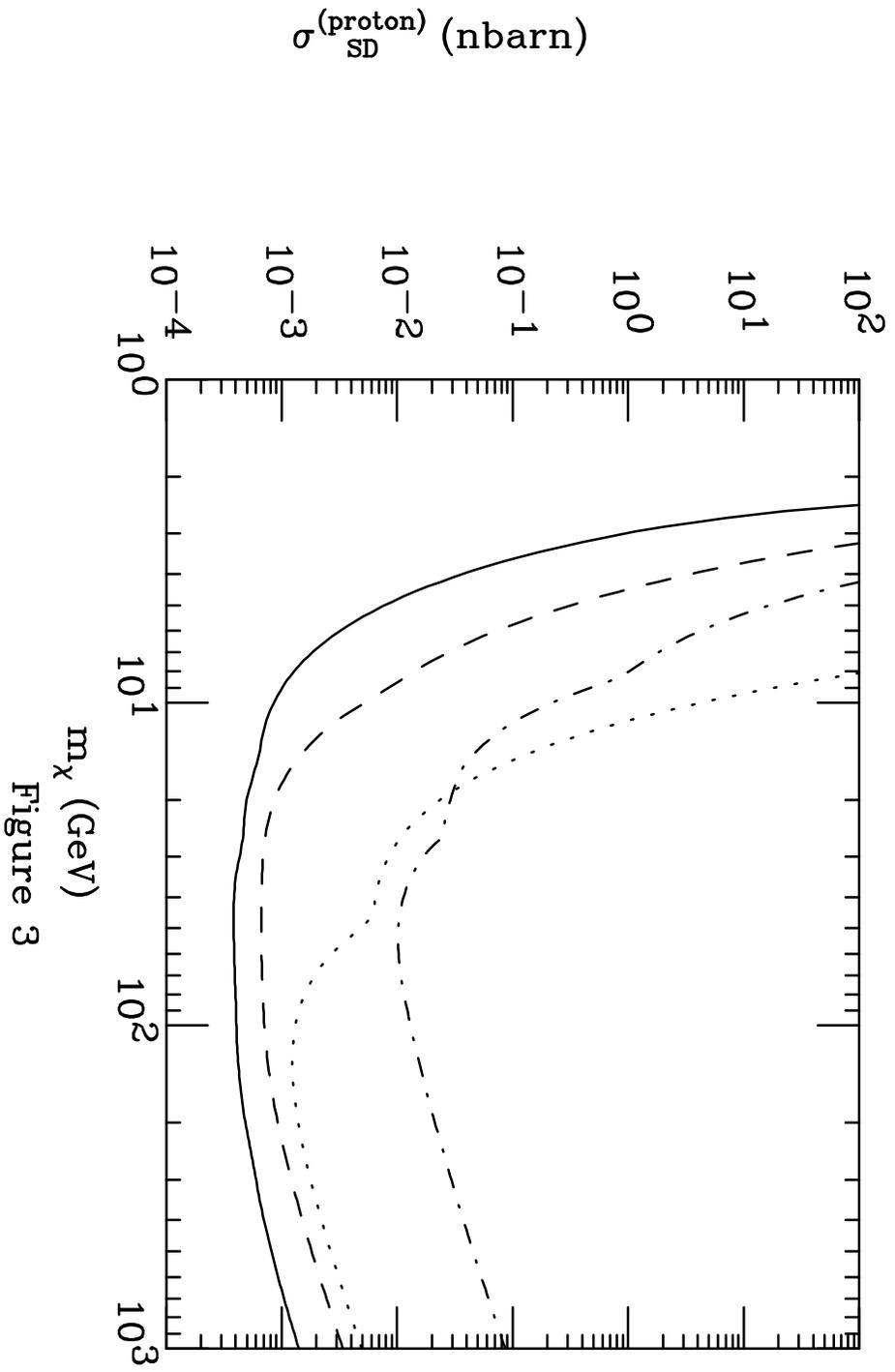
{\bf Figure 3} -- Upper bounds on $\sigma_{SD}^{(proton)}$. 
The solid and the dashed lines denote the 
bounds from the results of NaI-detectors: the solid line is from data of 
Ref.\cite{dama}, the dashed line is from data of Ref.\cite{boulby}. 
The dotted 
line is from data of a Xe-detector \cite{rome} and the dot-dashed line is from 
Ge-detectors \cite{got,twin,cosme}.

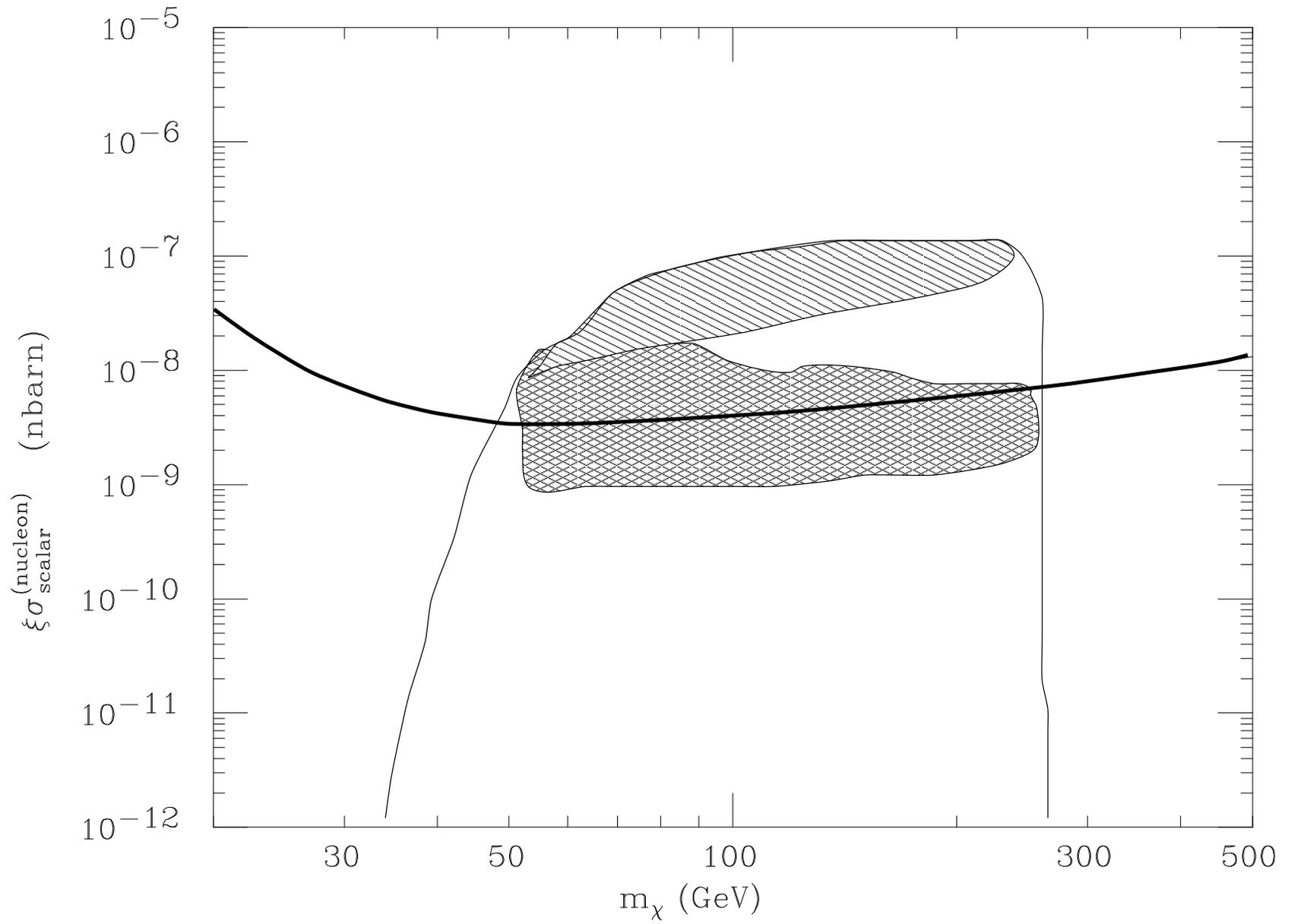
{\bf Figure 4}
  $\sigma^{(nucleon)}_{scalar}$, multiplied by the scaling factor $\xi$, versus
  the neutralino mass.
Allowed neutralino configurations stay inside the outer 
contour line. Configurations with 60 GeV $\leq m_A \leq$ 70 GeV and 
$\tan \beta = 50$ ($\tan \beta = 10$) stay in the up-left to bottom-right 
hatched region (in the doubly hatched region). 
The thick line denotes the upper bound obtained from data of Ref.\cite{dama}. 
The set I 
of Table I for the astrophysical and cosmological parameters has been used.

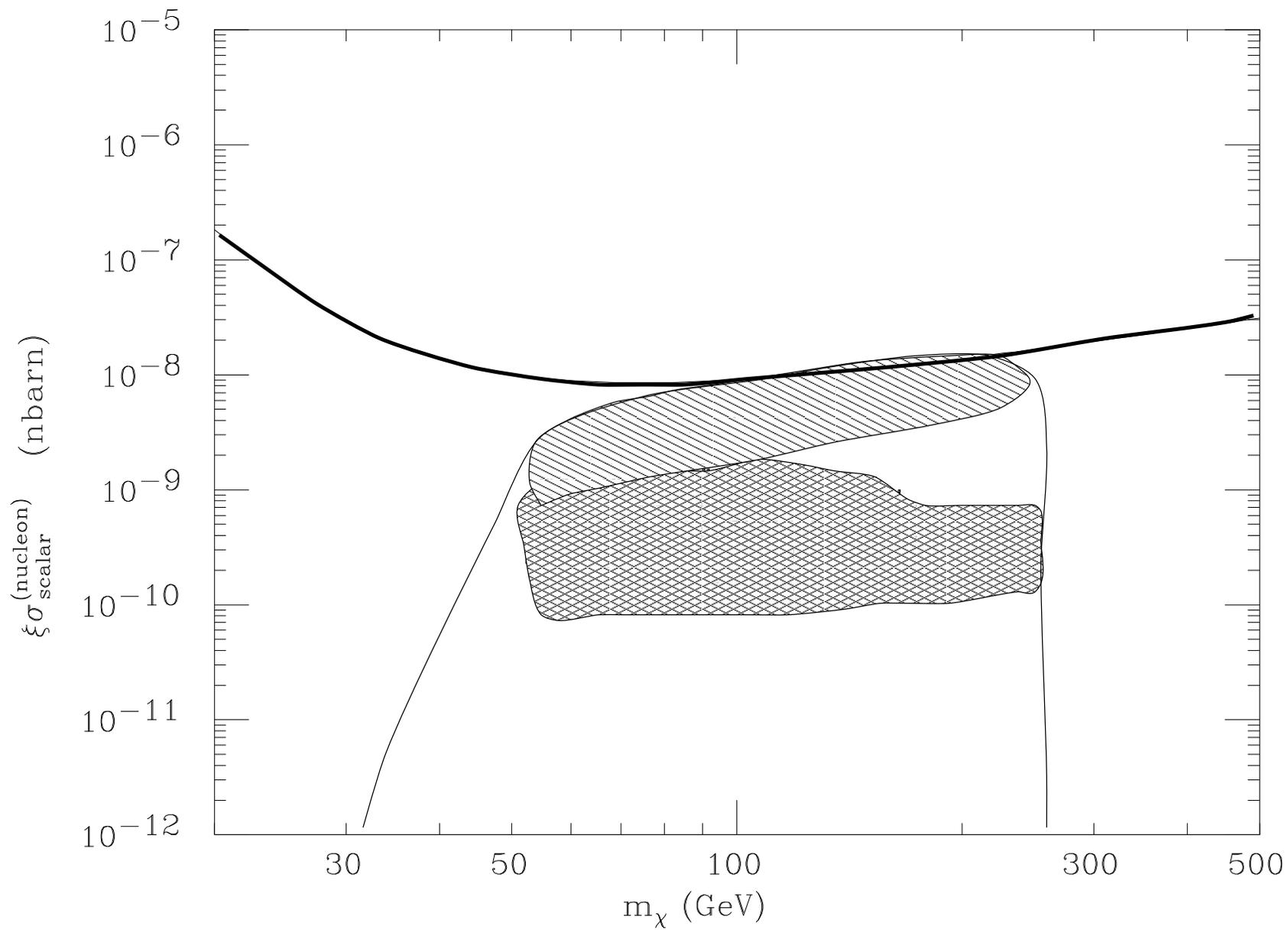
{\bf Figure 5}
The same as in Fig.4 except that set II
of Table I for the astrophysical and cosmological parameters has been used. 
\end{document}